\documentstyle[11pt]{article}
\normalbaselines
\newcounter{eqnletter}[equation]
\catcode`\@=11
\@addtoreset{equation}{section}   
\catcode`\@=12

\begin{document}

\begin{center}

{\LARGE\bf Bender--Dunne Orthogonal Polynomials \\[0.5cm]
And Quasi-Exact Solvability}

\vskip 1cm

{\large {\bf Agnieszka Krajewska\footnote{akraj@mvii.uni.lodz.pl}, 
             Alexander Ushveridze\footnote{alexush@mvii.uni.lodz.pl}}}\\
{\bf and}\\ 
{\large {\bf Zbigniew Walczak\footnote{walczak@mvii.uni.lodz.pl} }}

\vskip 0.1 cm

Department of Theoretical Physics, University of Lodz,\\
Pomorska 149/153, 90-236 Lodz, Poland

\end{center}
\vspace{1 cm}
\begin{abstract}

The paper is devoted to the further study of the remarkable classes of
orthogonal polynomials recently discovered by Bender and Dunne. We show
that these polynomials can be generated by solutions of arbitrary 
quasi - exactly solvable problems of quantum mechanichs both
one-dimensional and multi-dimensional.
A general and model-independent method for building and studying
Bender-Dunne polynomials is proposed. The method enables one to compute
the weight functions for the polynomials and shows that they are the
orthogonal polynomials in a discrete variable $E_k$ which takes its values
in the set of exactly computable energy levels of the corresponding
quasi-exactly solvable model.
It is also demonstrated that in an important particular case, the
Bender-Dunne polynomials exactly coincide with orthogonal polynomials
appearing in Lanczos tridiagonalization procedure.

\end{abstract}

\newpage

\section{Introduction}

Recently Bender and Dunne \cite{bender} 
showed that the solution $\psi(x,E)$ of the
Schr\"odinger equation $H\psi(x,E) = E\psi(x,E)$ for the one-dimensional
quasi-exactly solvable model with hamiltonian
$H = -\partial^2/\partial x^2 + c(c-1)x^{-2} + (4M+2c+3)x^2 +
x^6$ is the generating function for the set of polynomials 
$P_n(E)$ in the energy variable $E$. These polynomials,
defined as the coefficients of the expansion of function
$x^{-c}\exp(x^4/4)\psi(x,E)$ in powers of $x^2$, satisfy the
three-term recursion relations and therefore form an
orthogonal set with respect to some weight function $\omega(E)$.
Bender and Dunne demonstrated that a) for non-negative integer 
values of $M$ for which the model $H$ is quasi-exactly solvable, 
the norms of polynomials $P_n(E)$ vanish for $n\ge M+1$; b)
the zeros of the critical polynomial $P_{M+1}(E)$ coincide with
the exactly calculable energy levels in model $H$; c) the polynomials
$P_n(E)$ of degrees higher than $M+1$ factor into a product of 
two polynomials, one of which is $P_{M+1}(E)$:
$P_{M+1+n}(E) = P_{M+1}(E)Q_n(E)$; d) the factor-polynomials
$Q_n(E)$ also form an orthogonal set.

These and other remarkable properties of polynomials
$P_n(E)$ discussed in paper \cite{bender} suggest 
that they can be considered as an alternative mathematical
language for describing the phenomenon of quasi-exact solvability.
This phenomenon has been discovered several years ago 
(see e.g. the original papers \cite{zasl, turush, ush, tur}
and books \cite{kam, ushbook}).
In their paper, Bender and Dunne did not present any proof 
of the existence of polynomials $P_n(E)$ with similar properties 
for other quasi-exactly solvable models. Many important questions,
such as the problem of construction of some analogues of
Bender -- Dunne polynomials for multi-dimensional
quasi-exactly solvable models, the problem of the
computation of weight functions associated with polynomials
$P_n(E)$ and the corresponding factor-polynomials $Q_n(E)$,
also remained to be open.

The aim of the present paper is threefold. First of all, we intend
to propose a model independent way of introducing Bender --
Dunne polynomials for quasi-exactly solvable systems
of quantum mechanics. We show that the polynomials $P_n(E)$
(which we call the Bender -- Dunne polynomials of genus
one) and the corresponding factor-polynomials $Q_n(E)$ (which we
call the Bender -- Dunne polynomials of genus two) naturally appear 
as the coefficients of the expansion of solutions $\psi(x,E)$ of
Schr\"odinger equation in the bases in which the
hamiltonian $H$ has a tridiagonal form 
\footnote{Note that the hamiltonians of all quasi-exactly solvable
models are explicitly tridiagonalizable (for details see section
7 -- 9)}. 
Second, we present a simple way for building the weight functions 
$\omega(E)$ and $\rho(E)$ associated with polynomials $P_n(E)$ and ]
$Q_n(E)$ and show that these polynomials are orthogonal polynomials
of a discrete variable $E_k$. In the case of $P_n(E)$
polynomials, the variable $E_k$ takes its values in 
a finite set of the exactly calculable energy
levels in model $H$, and, in the case of $Q_n(E)$
polynomials, it runs an infinite set of all the remaining
(exactly non-calculable) levels.
Third, we demonstrate that the Bender -- Dunne polynomials exactly
coincide with orthogonal polynomials in an operator
variable $H$ arising in the so-called Lanczos
tridiagonalization procedure \cite{lanczos, wilkinson, ushpt}. 
Since this procedure is
applicable to any quasi-exactly solvable models (both
one-dimensional and multi-dimensional), this enables one to
make a conclusion of the universality of Bender -- Dunne construction.

\section{Bender -- Dunne polynomials of genus one}

Consider a hamiltonian $H$ acting in Hilbert space $W$ in
which it has an infinite and discrete spectrum $E_k,\
k=0,1,\ldots, \infty$. Let $\phi_n,\ n=0,1,\ldots,\infty$
be a certain (not neccessarily orthogonal) basis in $W$ in
which the hamiltonian takes an explicit tridiagonal form.
This means that
\begin{eqnarray}
H\phi_n=A_n\phi_{n-1}+B_n\phi_n+C_n\phi_{n+1}, \quad n=0,1,\ldots,\infty
\label{2.1}
\end{eqnarray}
where $A_n$, $B_n$ and $C_n$ are some algebraically computable coefficients.
Note that values of these coefficients depened also on the normalization 
of basis vectors $\phi_n$.

Assume that there exists a certain non-negative integer $M$
for which
\begin{eqnarray}
C_M=0, \quad C_n \neq 0, \quad \mbox{for}\ n \neq M.
\label{2.2}
\end{eqnarray}
Then, the model with hamiltonian $H$ becomes quasi-exactly
solvable. Indeed,  from formulas (\ref{2.1}) and (\ref{2.2}) it follows
that a linear hull of the basis vectors
$\phi_0,\ldots,\phi_M$ forms a $(M+1)$-dimensional invariant subspace
$W_M\subset W$ for the hamiltonian $H$.
For this reason, the spectral problem for $H$ in $W$ breaks up into two
independent spectral problems, one of which is formulated
in the $(M+1)$-dimensional space $W_M$. This enables one to determine 
at least $M+1$ eigenvalues $E_0, E_1, \ldots, E_M$ of hamiltonian $H$ 
in a purely algebraic way. The remaining eigenvalues $E_{M+1},E_{M+2},
\ldots$ cannot be determined algebraically and therefore
the model under consideration is quasi-exactly solvable 
(see ref.\cite{ushbook}).
Hereafter we shall call the levels $E_0,\ldots, E_M$
exactly calculable and the levels $E_{M+1},E_{M+2},\ldots$
-- exactly non-calculable.

Let us now consider the Schr\"odinger equation
\begin{eqnarray}
H\psi(E)=E\psi(E)
\label{2.3}
\end{eqnarray}
in which $E$ is an arbitrary parameter.We look for its
solution in the form of the formal expansion
\begin{eqnarray}
\psi(E)=\sum_{n=0}^\infty \phi_n P_n(E)
\label{2.4}
\end{eqnarray}
in which $P_n(E)$ denote certain functions of $E$.

Note that the expansion (\ref{2.4}) diverges for $E\neq E_k$ 
and converges for $E=E_k$. If the model is
quasi-exactly solvable, this is, if the condition (\ref{2.2})
is satisfied, then the series (\ref{2.4}) becomes finite for
all $E=E_k,\ k=0,\ldots,M$. This means that for these
values of $E$ the values of functions $P_n(E)$ should vanish
for $n\ge M+1$:
\begin{eqnarray}
P_n(E_k)=0, \quad n\ge M+1,\quad k=0,1,\ldots, M.
\label{2.5}
\end{eqnarray}

In order to determine the form of functions $P_n(E)$ it is
sufficient to substitute the expansion (\ref{2.4}) into
equation (\ref{2.3}) and use the tridiagonality condition (\ref{2.1}).
This gives the recurrence relations for $P_n(E)$,
\begin{eqnarray}
EP_n(E)=A_{n+1}P_{n+1}(E)+B_nP_n(E)+C_{n-1}P_{n-1}(E),\quad 
n=0,1,\ldots,\infty,
\label{2.8}
\end{eqnarray}
which can be supplemented by the initial conditions
\begin{eqnarray}
P_{-1}(E)=0,\quad P_0(E)=1.
\label{2.9}
\end{eqnarray}

From formulas (\ref{2.8}) and (\ref{2.9}) it immediately
follows that functions $P_n(E)$ are polynomials of degree $n$.
From the general theory of orthogonal polynomials 
(see e.g. \cite{szego})
we know that $P_n(E)$ are orthogonal polynomials. This means that
there exists a certain weight function $\omega(E)$, which
can be normalized as
\begin{eqnarray}
\int \omega(E)dE=1, 
\label{2.10}
\end{eqnarray} 
for which
\begin{eqnarray}
\int P_n(E)P_m(E)\omega(E)dE=p_np_m\delta_{nm},
\label{2.11}
\end{eqnarray} 
where $p_n$ denote the norms of polynomials $P_n(E)$.
These norms can be found from the recurrence relations
(\ref{2.8}) after multiplying them by $E^{n-1}\omega(E)$ and
taking integral over $E$. The result is
\begin{eqnarray}
p_n=\left|\prod_{m=1}^n \frac{C_{m-1}}{A_m}\right|^{\frac{1}{2}}.
\label{2.13}
\end{eqnarray} 
If the model is quasi-exactly solvable, then from the condition (\ref{2.2})
and formula (\ref{2.13}) it follows that
\begin{eqnarray}
p_n=0,\quad n\ge M+1,
\label{2.15}
\end{eqnarray} 
i.e. the norms of all polynomials $P_n(E)$ with $n\ge M+1$ vanish. 
Hereafter, we shall call the polynomials $P_n(E)$ the
Bender -- Dunne polynomials of genus one.

\section{Factorization property}

The $P_n(E)$ polynomials of degrees $n\ge M+1$ have a 
remarkable factorization property first noted by Bender and
Dunne \cite{bender}. Indeed, 
from formula (\ref{2.5}) it follows that 
the $M+1$ exactly calculable eigenvalues $E_0,E_1,\ldots, E_M$ of 
hamiltonian $H$ are the roots of these polynomials.
This enables one to write
\begin{eqnarray}
P_{M+1}(E)\sim\prod_{n=0}^M (E-E_n)
\label{2.16}
\end{eqnarray} 
and
\begin{eqnarray}
P_{M+1+n}(E)=P_{M+1}(E)Q_n(E),
\label{2.17}
\end{eqnarray} 
where $Q_n(E)$ are certain
polynomials of degrees $n=0,1,\ldots,\infty$.

The first formula (\ref{2.16}) gives a practical way of determining
the exactly calculable eigenvalues of the hamiltonian $H$
from the equation 
\begin{eqnarray}
P_{M+1}(E)=0.
\label{2.18}
\end{eqnarray}  
The second formula (\ref{2.17}) expresses an important factorizability
property of the polynomials $P_n(E)$ with $n\ge M+1$.
The factor-polynomials $Q_n(E)$ we shall call the Bender --
Dunne polynomials of genus two.

\section{Bender -- Dunne polynomials of genus two}

Substituting (\ref{2.17}) into (\ref{2.8}), one can obtain the
recurrence relations immediately for the factor-polynomials
$Q_n(E)$:
\begin{eqnarray}
EQ_n(E)=A_{M+2+n}Q_{n+1}(E)+B_{M+1+n}Q_n(E)+C_{M+n}Q_{n-1}(E),
\quad n=0,1,\ldots,\infty.
\label{2.19}
\end{eqnarray}
Since $C_M=0$, it is sufficient to supplement the relations (\ref{2.19})
by the initial conditions
\begin{eqnarray}
Q_0(E)=1.
\label{2.20}
\end{eqnarray}
We see that $Q_n(E)$ are again the orthogonal polynomials. Denoting
the corresponding weight function by $\rho(E)$ and
normalizing it as  
\begin{eqnarray}
\int \rho(E)dE=1, 
\label{2.21}
\end{eqnarray} 
we can write
\begin{eqnarray}
\int Q_n(E)Q_m(E)\rho(E)dE=q_nq_m\delta_{nm}.
\label{2.22}
\end{eqnarray} 
The norms of polynomials $Q_n(E)$ computed from the recurrence relations
(\ref{2.19}) are
\begin{eqnarray}
q_n=\left|\prod_{m=1}^n \frac{C_{M+m}}{A_{M+m+1}}\right|^\frac{1}{2}.
\label{2.24}
\end{eqnarray} 
According to (\ref{2.2}), the norms of $Q_n(E)$ polynomials do not vanish.

\section{Weight functions}

In this section we compute the weight functions $\omega(E)$
and $\rho(E)$ for Bender -- Dunne orthogonal polynomials of
genus one and two. 

\medskip
{\bf Weight function for $P_n(E)$ polynomials.} From formulas
(\ref{2.9}) -- (\ref{2.11}) it follows that 
\begin{eqnarray}
\int P_n(E)\omega(E)dE=\delta_{n0}.
\label{5.1}
\end{eqnarray} 
Multiplying the relation (\ref{2.4}) by $\omega(E)$, 
taking the integral over $E$ and using (\ref{5.1}), we obtain
\begin{eqnarray}
\int \psi(E)\omega(E)dE=\phi_0.
\label{5.2}
\end{eqnarray} 
The basis element $\phi_0$ can obviously be represented as
a linear combination of eigenvectors of hamiltonian $H$
corresponding to exactly calculable energy levels:
\begin{eqnarray}
\phi_0=\sum_{k=0}^M \omega_k\psi(E_k).
\label{5.3}
\end{eqnarray} 
Substituting (\ref{5.3}) into (\ref{5.2}), we obtain
\begin{eqnarray}
\int \psi(E')\omega(E')dE'=\sum_{k=0}^M \omega_k\psi(E_k).
\label{5.4}
\end{eqnarray} 
Let us now act on both hand sides of the relation (\ref{5.4})
by the operator function $\delta(H-E)$. Using (\ref{2.3}),
we obtain
\begin{eqnarray}
\psi(E)\omega(E)=\sum_{k=0}^M \omega_k\delta(E-E_k)\psi(E_k).
\label{5.6}
\end{eqnarray} 
Dividing both hand sides of (\ref{5.6}) by $\psi(E)$ and
taking into account the projection property of
delta-function, we obtain the final expression for the
weight function $\omega(E)$:
\begin{eqnarray}
\omega(E)=\sum_{k=0}^M \omega_k\delta(E-E_k).
\label{5.7}
\end{eqnarray} 
Thus, we have arrived at interesting result: the Bender --
Dunne polynomials of genus one are orthogonal polynomials
in a discrete variable $E_k, \ k=0,1,\ldots, M$ which takes
its values in the finite set of the exactly calculable eigenvalues 
of hamiltonian $H$. The orthogonality property for these polynomials
can now be written in the form
\begin{eqnarray}
\sum_{k=0}^M \omega_k P_n(E_k)P_m(E_k)=p_np_m\delta_{nm}.
\label{5.8}
\end{eqnarray} 
The coefficients $\omega_k$ can be computed algebraically.
Indeed, substituting into (\ref{5.3}) the expansions
(\ref{2.4}), we obtain the system of algebraic equations 
for numbers $\omega_k$:
\begin{eqnarray}
\sum_{k=0}^M P_n(E_k)\omega_k=\delta_{n0}, \quad n=0,1,\ldots, M.
\label{5.9}
\end{eqnarray} 
The numerical analysis of this system for various models
shows that the numbers $\omega_k$, playing the role of a
discrete weight function are always positive (see e.g.
section 6). It would be
an interesting task to proof this in general case.

\medskip
{\bf Weight function for $Q_n(E)$ polynomials.} Using formulas
(\ref{2.20}) -- (\ref{2.22}) we obtain
\begin{eqnarray}
\int Q_n(E)\rho(E)dE=\delta_{n0}.
\label{5.10}
\end{eqnarray} 
Using the factorization property (\ref{2.17}), let us 
represent the expresion (\ref{2.4}) in the form
\begin{eqnarray}
\psi(E)=\sum_{n=0}^M\phi_nP_n(E)+P_{M+1}(E)\sum_{n=0}^\infty
\phi_{M+1+n}Q_n(E).
\label{5.11}
\end{eqnarray} 
Multiplying (\ref{5.11}) by $\rho(E)P_{M+1}^{-1}(E)$,
taking the integral over $E$ and using (\ref{5.10}), we obtain
\begin{eqnarray}
\int \frac{\psi(E')}{P_{M+1}(E')}\rho(E')dE'=
\sum_{m=0}^M\phi_m\int\frac{P_m(E')}{P_{M+1}(E')}
\rho(E')dE'+\phi_{M+1}.
\label{5.12}
\end{eqnarray} 
Expressing vectors $\phi_n,\ n=0,1,\ldots, M$  and $\phi_{M+1}$ 
via the orthonormalized eigenvectors of hamiltonian $H$, we can rewrite
(\ref{5.12}) as
\begin{eqnarray}
\int \frac{\psi(E')}{P_{M+1}(E')}\rho(E')dE'=
\sum_{k=0}^M\mu_k\psi(E_k)+\sum_{k=M+1}^\infty\nu_k\psi(E_k),
\label{5.13}
\end{eqnarray}  
where $\mu_k$ and $\nu_k$ are certain, in general, algebraically
non-computable coefficients. Note that 
\begin{eqnarray}
\nu_{k}=\langle\phi_{M+1},\psi(E_k)\rangle,\quad
k=M+1,\ldots,\infty,
\label{5.14}
\end{eqnarray}  
where $\langle\ ,\ \rangle$ denotes the scalar product in
Hilbert space $W$.
Let us now act on both hand sides of the relation (\ref{5.13})
by the operator function $\delta(H-E)$. Using (\ref{2.3}),
we obtain
\begin{eqnarray}
\frac{\psi(E)}{P_{M+1}(E)}\rho(E)=
\sum_{k=0}^M \mu_k\delta(E-E_k)\psi(E_k)+
\sum_{k=M+1}^\infty \nu_k\delta(E-E_k)\psi(E_k).
\label{5.16}
\end{eqnarray} 
Dividing both hand sides of (\ref{5.16}) by $\psi(E)P_{M+1}^{-1}(E)$ and
taking into account the projection property of
delta-function and formula (\ref{2.5}), 
we obtain the final expression for the
weight function $\rho(E)$:
\begin{eqnarray}
\rho(E)=\sum_{k=M+1}^\infty \rho_k\delta(E-E_k),
\label{5.17}
\end{eqnarray}
where  
\begin{eqnarray}
\rho_k=P_{M+1}(E_k)\nu_{k}=
P_{M+1}(E_k)\langle\phi_{M+1},\psi(E_k)\rangle,\quad
k=M+1,\ldots,\infty.
\label{5.18}
\end{eqnarray}  
Thus, we see that the Bender --
Dunne polynomials of genus two are the orthogonal polynomials
in a discrete variable $E_k, \ k=M+1,\ldots, \infty$ whose values
are nothing else than the exactly non-calculable eigenvalues 
of hamiltonian $H$. The orthogonality property for these polynomials
can now be written in the form
\begin{eqnarray}
\sum_{k=M+1}^\infty \rho_k Q_n(E_k)Q_m(E_k)=q_nq_m\delta_{nm}.
\label{5.19}
\end{eqnarray} 
Note that the numbers $\rho_k$ cannot be found by means of algebraic
methods. 

\section{An example}

In this section we construct the discrete weight function $\omega_k$ 
for Bender -- Dunne polynomials of genus one associated
with the simplest quasi-exactly solvable models with
polynomial potentials.
The hamiltonians of these models, which were first discussed
in ref. \cite{turush}, have the form
\begin{eqnarray}
H=-\frac{\partial^2}{\partial x^2}+x^6+2bx^4+[b^2-4M-3]x^2,
\label{7.1}
\end{eqnarray}
and for any given non-negative integer $M$ admit $M+1$
explicit solutions corresponding to energy levels $E_0,
E_1,\ldots, E_M$ describing the first $M+1$ even states
with ordinal numbers $0,2,\ldots, 2M$.
It is easily seen that in the basis
\begin{eqnarray}
\phi_n(x)=\frac{(-x^2)^n}{(2n)!} 
\exp\left[-\frac{x^4}{4}-\frac{bx^2}{2}\right]
\label{7.2}
\end{eqnarray}
the hamiltonian $H$ takes an explicit tridiagonal form.
The corresponding coefficients $A_n$, $B_n$ and $C_n$ are
\begin{eqnarray}
A_n=1, \quad B_n=b(4n+1),\quad C_n=8a(n+1)(2n+1)(n-M).
\label{7.3}
\end{eqnarray}
The general formulas of sections 2 -- 5 allow one to
reconstruct the polynomials $P_n(E)$ and compute the
corresponding discrete weight functions.
Below we consider two examples of such calculations.

\medskip
{\bf Example 1.} Let us take $b=0$ and $M=8$. Then Bender
-- Dunne polynomials of genus one reconstructed from
recurrence relations read
\begin{eqnarray}
P_{0}(E)&=& 1, \nonumber\\
P_{1}(E)&=& E, \nonumber\\
P_{2}(E)&=& -64 + {E^2}, \nonumber\\
P_{3}(E)&=& -400\,E + {E^3}, \nonumber\\
P_{4}(E)&=& 46080 - 1120\,{E^2} + {E^4}, \nonumber\\
P_{5}(E)&=& 494080\,E - 2240\,{E^3} + {E^5}, \nonumber\\
P_{6}(E)&=& -66355200 + 2106880\,{E^2} - 3680\,{E^4} + 
                                         {E^6}, \nonumber\\
P_{7}(E)&=& -848977920\,E + 5655040\,{E^3} - 5264\,{E^5} +
                                                 {E^7}, \nonumber\\
P_{8}(E)&=& 96613171200 - 3916595200\,{E^2} + 11013120\,{E^4} -
6720\,{E^6} + {E^8}, \nonumber\\
P_{9}(E)&=& 911631974400\,E - 9345433600\,{E^3} + 16066560\,{E^5}
- 7680\,{E^7} + {E^9}. 
\label{7.4}
\end{eqnarray}
The energy levels coinciding with the roots of the last
polynomial $P_{9}(E)$ are
\begin{eqnarray}
E_{0}&=& -68.1568, \nonumber\\
E_{1}&=& -46.5609, \nonumber\\
E_{2}&=& -27.2997, \nonumber\\
E_{3}&=& -11.021, \nonumber\\
E_{4}&=& 0, \nonumber\\
E_{5}&=& 11.021, \nonumber\\
E_{6}&=& 27.2997, \nonumber\\
E_{7}&=& 46.5609, \nonumber\\
E_{8}&=& 68.1568, 
\label{7.5}
\end{eqnarray}
and the components of the corresponding weight function $\omega_k$
computed by means of formula (\ref{5.9}) are
\begin{eqnarray}
\omega_{0}&=& 0.0000117487, \nonumber\\
\omega_{1}&=& 0.00058638, \nonumber\\
\omega_{2}&=& 0.013045, \nonumber\\
\omega_{3}&=& 0.172498, \nonumber\\
\omega_{4}&=& 0.627717, \nonumber\\
\omega_{5}&=& 0.172498, \nonumber\\
\omega_{6}&=& 0.013045, \nonumber\\
\omega_{7}&=& 0.00058638, \nonumber\\
\omega_{8}&=& 0.0000117487. \nonumber
\label{7.6}
\end{eqnarray}
We see that all these numbers are positive. Note also that
these numbers are disctributed symmetrically with respect
to the reflection $k\rightarrow M+1-k$, while the energy
levels are distributed anti-symmetrically. This is a
trivial consequence of the self duality property of the
model (\ref{7.1}) with $b=0$ noted in ref. $\cite{aaz1}$.

\medskip
{\bf Example 2.} Let us now take $b=3$ and $M=8$. Then Bender
-- Dunne polynomials of genus one reconstructed from
recurrence relations read
\begin{eqnarray}
P_{0}(E)&=& 1, \nonumber\\
P_{1}(E)&=& -3 + E, \nonumber\\
P_{2}(E)&=& -19 - 18\,E + {E^2}, \nonumber\\
P_{3}(E)&=& 1521 + 131\,E - 45\,{E^2} + {E^3}, \nonumber\\
P_{4}(E)&=& -45639 + 9372\,E + 1166\,{E^2} - 84\,{E^3} +
{E^4}, \nonumber\\
P_{5}(E)&=& 624069 - 670331\,E + 306\,{E^2} + 4330\,{E^3} -
135\,{E^4} + {E^5}, \nonumber\\
P_{6}(E)&=& 26403813 + 29359242\,E - 2368649\,{E^2} - 151524\,{E^3} +
11395\,{E^4} +  \nonumber\\
&& - 198\,{E^5} + {E^6}, \nonumber\\
P_{7}(E)&=& -2968811271 - 1113735033\,E + 206523213\,{E^2} + 
   2136931\,{E^3} +  \nonumber\\
&& - 792309\,{E^4} + 24661\,{E^5} - 273\,{E^6} + {E^7}, \nonumber\\
P_{8}(E)&=& 219842628849 + 51179080248\,E - 15632501620\,{E^2} +
241229160\,{E^3} +  \nonumber\\
&& + 54476694\,{E^4} - 2649528\,{E^5} + 46956\,{E^6} - 360\,{E^7} +
{E^8}, \nonumber\\
P_{9}(E)&=& -18914361435891 - 3777700684023\,E +
1400534456148\,{E^2} +  \nonumber\\
&& - 41565642220\,{E^3} - 4391346906\,{E^4} +
293105406\,{E^5} - 7036092\,{E^6} +  \nonumber\\
&& + 81636\,{E^7} - 459\,{E^8} + {E^9}. 
\label{7.7}
\end{eqnarray}
For the energy levels we have
\begin{eqnarray}
E_{0}&=& -14.4717, \nonumber\\
E_{1}&=& -2.52391, \nonumber\\
E_{2}&=& 6.59852, \nonumber\\ 
E_{3}&=& 21.606, \nonumber\\
E_{4}&=& 40.6664, \nonumber\\
E_{5}&=& 62.6983, \nonumber\\
E_{6}&=& 87.2765, \nonumber\\
E_{7}&=& 114.122, \nonumber\\
E_{8}&=& 143.028, 
\label{7.8}
\end{eqnarray}
and the components of the weight function are
\begin{eqnarray}
\omega_{0}&=& 0.026295, \nonumber\\
\omega_{1}&=& 0.475485, \nonumber\\
\omega_{2}&=& 0.423177, \nonumber\\
\omega_{3}&=& 0.067025, \nonumber\\
\omega_{4}&=& 0.00741457, \nonumber\\
\omega_{5}&=& 0.000573463, \nonumber\\
\omega_{6}&=& 0.0000291886, \nonumber\\
\omega_{7}&=& 8.80537\; {{10}^{-7}}, \nonumber\\
\omega_{8}&=& 1.1954\; {{10}^{-8}}. 
\label{7.9}
\end{eqnarray}
We see that the discrete weight function is again positive.
At the same time, we have no symmetry of its components and
no anti-symmetry of the energy levels, because for $b\neq 0$ 
the model (\ref{7.1}) is not self-dual \cite{aaz1}.

\section{One-dimensional quasi-exactly solvable models}

The hamiltonians $H=-\partial^2/\partial x^2+V(x)$ of
one-dimensional quasi-exactly solvable models introduced in
paper \cite{ush} are
distinguished by the fact that for them it is possible to
find two functions $\lambda(x)$ and $\rho(x)$ for which the
ansatz
\begin{eqnarray}
\psi(x,E)=f(\lambda(x),E)\rho(x)
\label{1.4}
\end{eqnarray}
reduces the corresponding Schr\"odinger equation 
$H\psi(x,E)=E\psi(x,E)$ to the form
\begin{eqnarray}
\left[
R_3(\lambda)
\frac{\partial^2}{\partial \lambda^2}+
R_2(\lambda)
\frac{\partial}{\partial \lambda}+
R_1(\lambda)
\right]
f(\lambda,E)=Ef(\lambda,E)
\label{1.5}
\end{eqnarray}
with some polynomials $R_1(\lambda)$, $R_2(\lambda)$ and $R_3(\lambda)$ 
of degrees $m_1=1$, $m_2=2$ and $m_3\le 3$, respectively. The phenomenon of
quasi-exact solvability appears when
\begin{eqnarray}
M(M-1)\lim_{\lambda\rightarrow\infty}\lambda^{-3}R_3(\lambda)+
M\lim_{\lambda\rightarrow\infty}\lambda^{-2}R_2(\lambda)+
\lim_{\lambda\rightarrow\infty}\lambda^{-1}R_1(\lambda)=0,
\label{1.6}
\end{eqnarray}
where $M$ is a certain non-negative integer. 
It is easily seen that, for any $M$, there exist $M+1$ values of $E=E_k$
for which the equation (\ref{1.6}) has solutions
$f(\lambda,E)=f(\lambda,E_k)$ represented by certain
polynomials of degree $M$. 
From (\ref{1.6}) it immediately follows that in the basis
\begin{eqnarray}
\phi_n(x)=[\lambda(x)-r]^n\rho(x), \quad n=0,1,\ldots,\infty
\label{1.7}
\end{eqnarray}
in which $r$ denotes any of the roots of the polynomial $R_3(\lambda)$,
the hamiltonian of the model (irrespective of whether the
condition of quasi-exact solvability is satisfied or not) takes an
explicit tridiagonal form (\ref{2.1}) with trivially computable coefficients
$A_n$, $B_n$ and $C_n$. Indeed, introducing the new variable $t=\lambda-r$
and taking $f(\lambda,E) \equiv g(t,E)$, we obtain instead of (\ref{1.5}):
\begin{eqnarray}
\left [
(a_3 t^3 + b_3 t^2 + c_3 t) \frac{\partial^2}{\partial t^2} + 
(a_2 t^2 + b_2 t + c_2) \frac{\partial}{\partial t} +
(a_1 t + b_1) 
\right ]
g(t,E) = E g(t,E).
\label{1.8}
\end{eqnarray}
It is easily seen that the operator $L$ standing in the left-hand of the
equation (\ref{1.8}) is a linear combination of three
operators $L^+$, $L^0$, $L^-$ acting on the monomials $t^n$ as $L^+ t^n 
\sim t^{n+1}$, $L^0 t^n \sim t^n$ and $L^- t^n \sim t^{n-1}$. But this
means that this $L$-operator is tridiagonal in the basis $t^n$, $n=0,\ldots,
\infty$. This, in turn, results in the tridiagonality of $H$ in basis
(\ref{1.7}). According to general prescriptions given above, this means
that any one-dimensional quasi-exactly solvable model from
the list given in ref. \cite{ush} trivially generates
$m_3\le 3$ Bender -- Dunne polynomials of genus one and two.
An explicit construction of these polynomials and
computation of the corresponding weight functions will be
given in a separate publication.

\section{Lanczos tridiagonalization procedure}

There are infinitely many ways of reducing a given
hamiltonian $H$ to a tridiagonal form. One of the simplest
ways is based on the use of the so-called Lanczos tridiagonalization 
procedure (see e.g. refs.\cite{lanczos, wilkinson, ushpt}).
In terms of the Lanczos procedure, the properties of Bender
-- Dunne polynomials become especially transparent and simple.

Let $H$ be a hamiltonian and 
$\phi\in W$ be an arbitrary vector of Hilbert space. Consider
the sequence of vectors
\begin{eqnarray}
\phi_n=\bar P_n(H)\phi,\quad n=0,1,\ldots,\infty,
\label{6.1}
\end{eqnarray} 
in which $\bar P_n(H)$ denote certain polynomials in the operator
variable $H$.
Let us choose these polynomials from the condition of orthonormalizability
of vectors $\phi_n$
\begin{eqnarray}
\langle \phi_n,\phi_m\rangle=\delta_{nm}
\label{6.2}
\end{eqnarray} 
with respect to the scalar product in the space $W$. If the
hamiltonian is given and the scalar product is explicitly defined,
then the orthogonalization is a purely algebraic procedure.
The sequence $\phi_n$ can be considered as an orthogonal
basis in Hilbert space. 
Let $\psi(E_k)$ be the orthonormalized eigenvectors of the
hamiltonian $H$,
\begin{eqnarray}
\langle \psi(E_k),\psi(E_l)\rangle=\delta_{kl}.
\label{6.3}
\end{eqnarray} 
Expanding the vector $\phi$ in eigenvectors $\psi(E_k)$,
\begin{eqnarray}
\phi=\sum_{k=0}^\infty c_k \psi(E_k),
\label{6.4}
\end{eqnarray} 
and substituting this expansion into (\ref{6.1}) we obtain
\begin{eqnarray}
\phi_n=\sum_{k=0}^\infty c_k P_n(E_k)\psi(E_k).
\label{6.5}
\end{eqnarray} 
The substitution of (\ref{6.5}) into the orthonormalizability
condition (\ref{6.2}) gives the relation
\begin{eqnarray}
\sum_{k=0}^\infty c_k^2\bar P_n(E_k)\bar P_m(E_k)=\delta_{nm}
\label{6.6}
\end{eqnarray} 
which shows that $\bar P_n(E)$ are orthogonal
polynomials in a discrete variable $E_k,\
k=0,1,\ldots,\infty$, which takes its values in the set of
eigenvalues of hamiltonian $H$ (see ref.\cite{ushpt}). 
The corresponding discrete weight function
$c_k^2$ is obviously positive. Hereafter, we shall call
$\bar P_n(H)$ the Lanczos polynomials.
Because of the orthogonality, the polynomials $\bar P_n(H)$ obey
the recurrence relation
\begin{eqnarray}
H\bar P_n(H)=A_n\bar P_{n-1}(H)+B_n\bar P_n(H)+C_n\bar P_{n+1}(H),
\quad n=0,1,\ldots,\infty,
\label{6.7}
\end{eqnarray} 
in which $A_n$, $B_n$ and $C_n$ are certain algebraically
computable coefficients.
Acting by the operator equality (\ref{6.7}) on $\phi$ we obtain
\begin{eqnarray}
H\phi_n=A_n\phi_{n-1}+B_n\phi_n+C_n\phi_{n+1},\quad n=0,1,\ldots,\infty,
\label{6.8}
\end{eqnarray} 
which means that the operator $H$ is tridiagonal in the basis
$\phi_n,\ n=0,1,\ldots,\infty$. Since $H$ is a hermitian operator and the
basis $\phi_n$ is orthogonal, the upper and lower diagonals of $H$ should
coincide. This gives us the conditions
\begin{eqnarray}
C_n=A_{n+1}
\label{6.9}
\end{eqnarray}
which reduce the recurrence relations (\ref{6.7}) to the form
\begin{eqnarray}
H\bar P_n(H)=C_{n-1}\bar P_{n-1}(H)+B_n\bar P_n(H)+C_n\bar P_{n+1}(H),
\quad n=0,1,\ldots,\infty.
\label{6.10}
\end{eqnarray}
Let us now consider the equation (\ref{2.3}) for the
hamiltonian $H$. Looking for its solutions in the form (\ref{2.4}),
substituting (\ref{2.4}) into (\ref{6.8}) and using formula
(\ref{6.9}) we find that
Bender -- Dunne polynomials $P_n$ associated with the expansion (\ref{6.8})
satisfy the recurrence relation
\begin{eqnarray}
EP_n(E)=C_{n-1}P_{n-1}(E)+B_nP_n(E)+C_n P_{n+1}(E),
\quad n=0,1,\ldots,\infty.
\label{6.12}
\end{eqnarray}
From the coincidence of the recurrence relations
(\ref{6.10}) and (\ref{6.12}) it follows that 
the Bender -- Dunne polynomials
$P_n(E)$ are nothing else that the Lanczos polynomials
$\bar P_n(E)$. Hereafter, we shall not distinguish between
the Lanczos and Bender -- Dunne polynomials and for both of
them use the same notation $P_n(E)$.

Up to now, we implicitly assumed that all the coefficients $c_k$
in the expansion (\ref{6.4}) differ from zero. Assume now
that this expansion has only a finite number of terms 
\begin{eqnarray}
\phi=\sum_{k=0}^M c_k \psi(E_k).
\label{6.13}
\end{eqnarray} 
Let us denote by $W_M$ the linear span of functions
$\psi(E_k), \ k=0,1,\ldots, M$. The space $W_M$ is a
$(M+1)$-dimensional invariant subspace for the hamiltonian
$H$. It is quite obvious that, in this case, the 
Lanczos procedure becomes finite and consists only of $M+1$
essential steps.
Indeed, assume that the Lanczos polynomials $P_n(E)$ with $0\le
n\le M$ are already known and we want to construct
the next polynomials $P_{M+1+n}(H)$. According to general
prescriptions of Lanczos theory, the vectors
$\phi_{M+1+n}$ should have the form $\phi_{M+1+n} = P_{M+1+n}(H)\phi$ 
and thus should belong to the space $W_M$. On the other hand, they
should be orthogonal to all the linearly independent basis vectors
$\phi_0,\phi_1,\ldots,\phi_M$ of the space $W_M$, which
means that $\phi_{M+1+n}=P_{M+1+n}(H)\phi=0$. This is possible
only if $P_{M+1+n}(H)=\prod_{k=0}^M (H-E_k)Q_n(H)$,
where $E_0, E_1, \ldots, E_M$ are the eigenvalues of the
hamiltonian $H$ in the space $W_M$ and $Q_n(H)$ are the
{\it arbitrary} polynomials of degrees $n=0,1,\ldots, \infty$.
The fact that the functions $\phi_{M+1+n}$ are non-normalizable,
prevents one from determining the polynomials $Q_n(H)$ uniquely.

The only way to construct the polynomials $Q_n(H)$ is to choose a new vector
\begin{eqnarray}
\phi_{M+1}=P_{M+1}(H)\sum_{k=M+1}^\infty c_k \psi(E_k)
\label{6.14}
\end{eqnarray} 
lying outside the space $W_M$ and apply to it 
the Lanczos tridiagonalization procedure.
This vector is, obviously, normalizable, and starting with
it, we can construct the set of orthonormalized vectors
$\phi_{M+1+n} = Q_n(H)\phi_{M+1}$ belonging to an orthogonal
complement of the space $W_M$. This procedure enables one to
determine the polynomials $Q_n(H)$ uniquely.

It is easily seen that the
orthonormalizability condition for functions $\phi_n, \
n=0,1,\ldots, M$ reads
\begin{eqnarray}
\sum_{k=0}^M c_k^2 P_n(E_k) P_m(E_k)=\delta_{nm}
\label{6.15}
\end{eqnarray} 
and is nothing else than the orthonormalizability condition
for Bender -- Dunne polynomials of genus one. Note the positive 
definiteness of the corresponding weight function $\omega_k=c^2_k$.
At the same time, the orthonormalizability condition for functions 
$\phi_{M+1+n}, \ n=0,1,\ldots, \infty$ takes the form
\begin{eqnarray}
\sum_{k=0}^M c_k^2 P_{M+1}^2(E_k)Q_n(E_k) Q_m(E_k)=\delta_{nm}
\label{6.16}
\end{eqnarray} 
and becomes that for Bender -- Dunne polynomials of genus two. 
The corresponding 
weight function $\rho_k=c^2_kP_{M+1}^2(E_k)$ is also
positively defined.

\section{Conclusion}

In conclusion note that the Lanczos tridiagonalization procedure
can be applied to any hermitean operators in Hilbert space.
This means that $H$ may be the hamiltonian of an arbitrary
one or multi-dimensional model. However, if we want to
construct the Bender -- Dunne polynomials of genus one, one
should take care that the trial function $\phi$ belongs to
a finite-dimensional invariant subspace $W_M$. This can
always be done for quasi-exactly solvable models of quantum
mechanics.  Indeed, consider a quantum model whose hamiltonian can
be represented in the form
\begin{eqnarray}
H=\sum_{\alpha,\beta} C_{\alpha\beta} J_\alpha(h)J_\beta(h)+
\sum_\alpha C_\alpha J_\alpha(h)
\label{6.17}
\end{eqnarray} 
where $C_{\alpha\beta}$ and $C_\alpha$ are some numbers and
$J_\alpha(h)$ are generators of a certain Lie algebra
realising a representation with highest weight $h$ and
having the form of first-order differential operators. 
In this case, it is natural to identify the vector $\phi$ with the
highest weight vector $|h\rangle$. If the representation is
infinite-dimensional and irreducible, then all the terms of
the Lanczos sequence $\phi_n=P_n(H)|h\rangle$ are linearly independent
and this sequence is infinite. Otherwise, the operator $H$
would have a finite-dimensional invariant subspace, which
would contradict to the condition of irreducibility of the representation.
Assume now that the representation is finite-dimensional,
so that the model (\ref{6.17}) is quasi-exactly solvable \cite{tur}.
Since the highest weight vector $|h\rangle$ belongs, by
definition, to the representation space, the 
Lanczos sequence will contain in this case only a finite number of
linearly independent terms. According to general reasonings
given above this naturally leads us to Bender -- Dunne
polynomials of genus one.

\end{document}